\documentclass[12pt,a4paper,twoside,english]{revtex4}
\usepackage[T1]{fontenc}
\usepackage[latin1]{inputenc}
\usepackage{graphicx}

\makeatletter

\usepackage{babel}
\makeatother

\begin{document}

\title{Plasmons and polaritons in a semi-infinite plasma and a plasma slab}

\author{{\normalsize M. Apostol and G. Vaman }}

\address{Department of Theoretical Physics, Institute of Atomic Physics, \\
Magurele-Bucharest Mg-6, POBox Mg-35, Romania \\
email: apoma@theory.nipne.ro}

\begin{abstract}
Plasmon and polariton modes are derived for an ideal semi-infinite
(half-space) plasma and an ideal plasma slab by using a general, unifying
procedure, based on equations of motion, Maxwell's equations and suitable
boundary conditions. Known results are re-obtained in much a more
direct manner and new ones are derived. The approach consists of representing
the charge disturbances by a displacement field in the positions of
the moving particles (electrons). The dielectric response and the
electron energy loss are computed. The surface contribution to the
energy loss exhibits an oscillatory behaviour in the transient regime
near the surfaces. The propagation of an electromagnetic wave in these
plasmas is treated by using the retarded electromagnetic potentials.
The resulting integral equations are solved and the reflected and
refracted waves are computed, as well as the reflection coefficient.
For the slab we compute also the transmitted wave and the transmission
coefficient. Generalized Fresnel's relations are thereby obtained
for any incidence angle and polarization. Bulk and surface plasmon-polariton
modes are identified. As it is well known, the field inside the plasma
is either damped (evanescent) or propagating (transparency regime),
and the reflection coefficient for a semi-infinite plasma exhibits
an abrupt enhancement on passing from the propagating regime to the
damped one (total reflection). Similarly, apart from characteristic
oscillations, the reflection and transmission coefficients for a plasma
slab exhibit an appreciable enhancement in the damped regime. 
\end{abstract}
\maketitle


\section{Introduction}

After the discovery of bulk plasmons in an infinite electron plasma,\citep{key-1}-\citep{key-3}
there was a great deal of interest in plasmons occurring in structures
with special geometries, like a half-space (semi-infinite) plasma,
a plasma slab of finite thickness, a two-plasmas interface (two plasmas
bounding each other), a two-dimesional sheet with an aperture, a slab
with a cilindrical hole, structures with surface gratings or regular
holes patterns, layered films, cilindrical rods and spherical particles,
etc. There is a vast literature on various structures with special
geometries exhibiting plasmon modes. These studies were aimed mainly
at identifying new plasmon modes, like the surface plasmons,\citep{key-4}-\citep{key-11}
accounting for the electron energy loss experiments and exploring
the interaction of the electron plasma with electromagnetic radiation
(polariton excitations).\citep{key-12}-\citep{key-24} More recently,
a possible enhancement of the electromagnetic radiation scattered
on electron plasmas with special geometries enjoyed a particular interest.\citep{key-25}-\citep{key-27}
In all these studies the plasmon and polariton modes are of fundamental
importance.\citep{key-28}-\citep{key-32} The methods used in deriving
such results are of great diversity, resorting often to particular
assumptions, such that the basic underlying mechanism of plasmons
or polaritons' occurrence is often obscured. The need is therefore
felt of having a general, unifying procedure for deriving plasmon
and polariton modes in structures with special geometries, as based
on the equation of motion of the charge density, Maxwell's equations
and the corresponding boundary conditions. Such a procedure is presented
in this paper for an ideal semi-infinite plasma and an ideal plasma
slab. 

We represent the charge disturbances as $\delta n=-ndiv\mathbf{u}$,
where $n$ is the (constant, uniform) charge concentration and $\mathbf{u}$
is a displacement field of the mobile charges (electrons). This representation
is valid for $\mathbf{K}\mathbf{u}(\mathbf{K})\ll1$, where $\mathbf{K}$
is the wavevector and $\mathbf{u}(\mathbf{K})$ is the Fourier component
of the displacement field. We assume a rigid neutralizing background
of positive charge, as in the well-known jellium model. In the static
limit, \emph{i.e.} for Coulomb interaction, the lagrangian of the
electrons can be written as \begin{equation}
L=\int d\mathbf{r}\left[\frac{1}{2}mn\dot{\mathbf{u}}^{2}-\frac{1}{2}\int d\mathbf{r}'U(\left|\mathbf{r}-\mathbf{r}'\right|)\delta n(\mathbf{r})\delta n(\mathbf{r}')\right]+e\int d\mathbf{r}\Phi(\mathbf{r})\delta n(\mathbf{r})\,\,\,,\label{1}\end{equation}
 where $m$ is the electron mass, $U(r)=e^{2}/r$ is the Coulomb energy,
$-e$ is the electron charge and $\Phi(\mathbf{r})$ is an external
scalar potential. Equation (\ref{1}) leads to the equation of motion
\begin{equation}
m\ddot{\mathbf{u}}=ngrad\int d\mathbf{r}'U(\left|\mathbf{r}-\mathbf{r}'\right|)div\mathbf{u}(\mathbf{r}')+egrad\Phi,\,\,\,\label{2}\end{equation}
which is the starting equation of our approach. We leave aside the
dissipation effects (which can easily be included in equation (\ref{2})). 

By using the Fourier transform for an infinite plasma it is easy to
see that the eigenmode of the homogeneous equation (\ref{2}) is the
well-known bulk plasmon mode given by $\omega_{p}^{2}=4\pi ne^{2}/m$.
On the other side, equation $\delta n=-ndiv\mathbf{u}$ is equivalent
with Maxwell's equation $div\mathbf{E}_{i}=-4\pi e\delta n$, where
$\mathbf{E}_{i}=4\pi ne\mathbf{u}$ is the internal electric field
(equal to $-4\pi\mathbf{P}$, where $\mathbf{P}$ is the polarization).
Making use of the electric displacement $\mathbf{D}=-grad\Phi=\varepsilon(\mathbf{D}+\mathbf{E}_{i})$,
we get the well-known dielectric function $\varepsilon=1-\omega_{p}^{2}/\omega^{2}$
in the long-wavelength limit from the solution of the inhomogeneous
equation (\ref{2}). Similarly, since the current density is $\mathbf{j}=-en\dot{\mathbf{u}}$,
we get the well-known electrical conductivity $\sigma=i\omega_{p}^{2}/4\pi\omega$.

We apply this approach to a semi-infinite plasma and a plasma slab.
First, we derive the surface and bulk plasmon modes and obtain the
dielectric response and the electron energy loss for a semi-infinite
plasma. The surface contribution to the energy loss exhibits an oscillatory
behaviour in the transient regime near the surface. Further on, we
consider the interaction of the semi-infinite plasma with the electromagnetic
field, as described by the usual term $(1/c)\int d\mathbf{r}\mathbf{j}\mathbf{A}-\int d\mathbf{r}\rho\Phi$
in the lagrangian, where $\mathbf{A}$ is the vector potential, $\rho=endiv\mathbf{u}$
is the charge density and $\Phi$ is the scalar potential. We limit
ourselves to the interaction with the electric field, and compute
the reflected and refracted waves, as well as the reflection coefficient.
Generalized Fresnel's relations are obtained for any incidence angle
and polarization. We find it more convenient to use the radiation
formulae for the retarded potentials, instead of using directly the
Maxwell's equations, and the resulting integral equations are solved.
Bulk and surface plasmon-polariton modes are identified. The field
inside the plasma is either damped (evanescent) or propagating (transparency
regime), and the reflection coefficient exhibits an abrupt enhancement
on passing from the propagating to the damping regime (total reflection).
Finally, we give similar results for a plasma slab, where we compute
also the transmitted field and the transmission coefficient. Apart
from characteristic oscillations, the reflection and transmission
coefficients for a plasma slab exhibit an appreciable enhancement
in the damped regime. The present approach can be extended to various
other plasma structures with special geometries.

\section{Plasma eigenmodes}

We consider an ideal semi-infinite plasma extending over the half-space
$z>0$ (and bounded by the vacuum for $z<0$). The displacement field
$\mathbf{u}$ is then represented as $(\mathbf{v},u_{3})\theta(z)$,
where $\mathbf{v}$ is the displacement component in the $(x,y)$-plane,
$u_{3}$ is the displacement component along the $z$-direction and
$\theta(z)=1$ for $z>0$ and $\theta(z)=0$ for $z<0$ is the step
function. In equation of motion (\ref{2}) $div\mathbf{u}$ is then
replaced by \begin{equation}
div\mathbf{u}=\left(div\mathbf{v}+\frac{\partial u_{3}}{\partial z}\right)\theta(z)+u_{3}(0)\delta(z)\,\,\,,\label{3}\end{equation}
 where $u_{3}(0)=u_{3}(\mathbf{r},z=0)$, $\mathbf{r}$ being the
in-plane ($x,y$) position vector. Equation (\ref{2}) becomes

\begin{equation}
\begin{array}{c}
m\ddot{\mathbf{u}}=ne^{2}grad\int d\mathbf{r}'dz'\frac{1}{\sqrt{(\mathbf{r}-\mathbf{r}')^{2}+(z-z')^{2}}}\left[div\mathbf{v}(\mathbf{r}'.z')+\frac{\partial u_{3}(\mathbf{r}',z')}{\partial z'}\right]+\\
\\+ne^{2}grad\int d\mathbf{r}'\frac{1}{\sqrt{(\mathbf{r}-\mathbf{r}')^{2}+z^{2}}}u_{3}(\mathbf{r}',0)+egrad\Phi\end{array}\label{4}\end{equation}
for $z>0$. One can see the (de)-polarizing field occurring at the
free surface $z=0$ (the second integral in equation (\ref{4})).

We use Fourier transforms of the type \begin{equation}
\mathbf{u}(r,z;t)=\sum_{\mathbf{k}}\int d\omega\mathbf{u}(\mathbf{k},z;\omega)e^{i\mathbf{kr}}e^{-i\omega t}\label{5}\end{equation}
 (for in-plane unit area), as well as the Fourier representation \begin{equation}
\frac{1}{\sqrt{r^{2}+z^{2}}}=\sum_{\mathbf{k}}\frac{2\pi}{k}e^{-k\left|z\right|}e^{i\mathbf{kr}}\label{6}\end{equation}
for the Coulomb potential. Then, it is easy to see that equation (\ref{4})
leads to the integral equation\begin{equation}
\omega^{2}v=\frac{1}{2}k\omega_{p}^{2}\int_{0}^{\infty}dz've^{-k\left|z-z'\right|}+\frac{1}{2k}\omega_{p}^{2}\int_{0}^{\infty}dz'\frac{\partial v}{\partial z^{'}}\frac{\partial}{\partial z^{'}}e^{-k\left|z-z'\right|}-\frac{iek}{m}\Phi\,\,\,\label{7}\end{equation}
and $iku_{3}=\frac{\partial v}{\partial z}$, where we have dropped
out for simplicity the arguments $\mathbf{k},\, z$ and $\omega$.
The $\mathbf{v}$-component of the displacement field is directed
along the wavevector $\mathbf{k}$ (in-plane longitudinal waves).
This integral equation can easily be solved. Integrating by parts
in its \emph{rhs} we get \begin{equation}
\omega^{2}v=\omega_{p}^{2}v-\frac{1}{2}\omega_{p}^{2}v_{0}e^{-kz}-\frac{iek}{m}\Phi\,\,\,,\label{8}\end{equation}
hence \begin{equation}
\begin{array}{c}
v=\frac{iek\omega_{p}^{2}}{m}\frac{\Phi_{0}}{(\omega^{2}-\omega_{p}^{2})(2\omega^{2}-\omega_{p}^{2})}e^{-kz}-\frac{iek}{m}\frac{\Phi}{\omega^{2}-\omega_{p}^{2}}\\
\\u_{3}=-\frac{ek\omega_{p}^{2}}{m}\frac{\Phi_{0}}{(\omega^{2}-\omega_{p}^{2})(2\omega^{2}-\omega_{p}^{2})}e^{-kz}-\frac{e}{m}\frac{\Phi^{'}}{\omega^{2}-\omega_{p}^{2}}\end{array}\label{9}\end{equation}
 where $v_{0}=v(z=0)$, $\Phi_{0}=\Phi(z=0)$ and $\Phi^{'}=\frac{\partial\Phi}{\partial z}$.
One can see the surface contributions (terms proportional to $\Phi_{0}e^{-kz}$)
and bulk contributions ($\Phi,\Phi^{'}$-terms). 

The solutions given by equations (\ref{9}) exhibit two eigenmodes,
the bulk plasmon $\omega_{b}=\omega_{p}$ and the surface plasmon
$\omega_{s}=\omega_{p}/\sqrt{2}$, as it is well known. Indeed, the
homogeneous equation (\ref{8}) ($\Phi=0$) has two solutions: the
surface plasmon $v=v_{0}e^{-kz}$ for $\omega^{2}=\omega_{p}^{2}/2$
and the bulk plasmon $v_{0}=0$ for $\omega^{2}=\omega_{p}^{2}$.
Making use of this observation we can represent the general solution
as an eigenmodes series \begin{equation}
v(\mathbf{k},z)=\sqrt{2k}v_{0}(\mathbf{k})e^{-kz}+\sum_{\kappa}\sqrt{\frac{2k^{2}}{\kappa^{2}+k^{2}}}v(\mathbf{k},\kappa)\sin\kappa z\,\,\,,\label{10}\end{equation}
for $z>0$, where $v(\mathbf{k},-\kappa)=-v(\mathbf{k},\kappa)$,
and $iku_{3}(\mathbf{k},z)=\frac{\partial v(\mathbf{k},z)}{\partial z}$.
Then, it is easy to see that the hamiltonian $H=T+U$ corresponding
to the lagrangian $L=T-U$ given by equation (\ref{1}) becomes \begin{equation}
\begin{array}{c}
T=nm\sum_{\mathbf{k}}\dot{v}_{0}^{*}(\mathbf{k})\dot{v}_{0}(\mathbf{k})+nm\sum_{\mathbf{k}\kappa}\dot{v}^{*}(\mathbf{k},\kappa)\dot{v}(\mathbf{k},\kappa)\\
\\U=2\pi n^{2}e^{2}\sum_{\mathbf{k}}v_{0}^{*}(\mathbf{k})v_{0}(\mathbf{k})+4\pi n^{2}e^{2}\sum_{\mathbf{k}\kappa}v^{*}(\mathbf{k},\kappa)v(\mathbf{k},\kappa)\,\,\,,\end{array}\label{11}\end{equation}
 where $T$ is the kinetic energy and $U$ is the potential energy.
We can see that this hamiltonian corresponds to harmonic oscillators
with frequencies $\omega_{s}=\omega_{p}/\sqrt{2}$ and $\omega_{b}=\omega_{p}$. 

Making use of $\mathbf{E}_{i}=4\pi ne\mathbf{u}$ and equations (\ref{9})
we can write down the internal field (polarization) as\begin{equation}
\begin{array}{c}
E_{\perp}(\mathbf{k},z;\omega)=\frac{ik\omega_{p}^{4}\Phi(\mathbf{k},0;\omega)}{(\omega^{2}-\omega_{p}^{2})(2\omega^{2}-\omega_{p}^{2})}e^{-kz}-\frac{ik\omega_{p}^{2}\Phi(\mathbf{k},z;\omega)}{\omega^{2}-\omega_{p}^{2}}\\
\\E_{\parallel}(\mathbf{k},z;\omega)=-\frac{k\omega_{p}^{4}\Phi(\mathbf{k},0;\omega)}{(\omega^{2}-\omega_{p}^{2})(2\omega^{2}-\omega_{p}^{2})}e^{-kz}-\frac{\omega_{p}^{2}\Phi^{'}(\mathbf{k},z;\omega)}{\omega^{2}-\omega_{p}^{2}}\end{array}\label{12}\end{equation}
where $E_{\perp}$ is directed along the in-plane wavevector $\mathbf{k}$
and $E_{\parallel}$ is parallel with the $z$-axis (perpendicular
to the surface $z=0$). This is the dielectric response of the semi-infinite
plasma to an external potential. 

We take an external potential of the form $\Phi(\mathbf{k},z)=\Phi^{0}(\mathbf{k})e^{i\kappa z}$
(leaving aside the frequency argument $\omega$), and get the electric
displacement $\mathbf{D}_{\perp}(\mathbf{k},z)=-i\mathbf{k}\Phi^{0}(\mathbf{k})e^{i\kappa z}$
and $D_{\parallel}(\mathbf{k},z)=-i\kappa\Phi^{0}(\mathbf{k})e^{i\kappa z}$
from $\mathbf{D}=-grad\Phi$. We can see that the surface terms do
not contribute to this response, as expected, since these terms are
localized. Making use of $\mathbf{E}_{i}=(1/\varepsilon-1)\mathbf{D}$,
we get the well-known dielectric function $\varepsilon(\kappa,\omega)=1-\omega_{p}^{2}/\omega^{2}$
in the long-wavelength limit.

\section{Electron energy loss}

It is well known that the energy loss per unit time (stopping power)
is given by \begin{equation}
P=\frac{d}{dt}\left(\frac{mv^{2}}{2}\right)=-e\mathbf{v}\mathbf{E}_{i}\,\,\,,\label{13}\end{equation}
 for an electron moving with velocity $\mathbf{v}=(\mathbf{v}_{\perp},v_{\parallel})$,
where the field $\mathbf{E}_{i}$ is taken at $\mathbf{r}=\mathbf{v_{\perp}}t$
and $z=v_{\parallel}t$ for $t>0$ ($z>0$). It is assumed that the
electron energy is sufficiently large and the energy loss is small
enough to use a constant $\mathbf{v}$ in estimating the \emph{rhs}
of equation (\ref{13}). The potential created by the electron is
given by the Poisson equation $\Delta\Phi=4\pi e\delta(\mathbf{r}-\mathbf{v}_{\perp}t)\delta(z-v_{\parallel}t)$,
whence, by making use of the Fourier representation (\ref{6}), we
get \begin{equation}
\Phi(\mathbf{k},z;\omega)=-\frac{2ev_{\parallel}}{(\omega-\mathbf{k}\mathbf{v}_{\perp})^{2}+k^{2}v_{\parallel}^{2}}e^{-i\left(\mathbf{k}\mathbf{v}_{\perp}-\omega\right)z/v_{\parallel}}\,\,.\label{14}\end{equation}
 We introduce this potential in equations (\ref{12}) and compute
the energy loss given by equation (\ref{13}). It contains two contributions,
one associated with the bulk plasmons, \begin{equation}
P_{b}=e^{2}\omega_{p}^{2}\sum_{\mathbf{k}}\int d\omega\frac{i\omega}{\omega_{p}^{2}-\omega^{2}}\cdot\frac{2v_{\parallel}}{(\omega-\mathbf{k}\mathbf{v}_{\perp})^{2}+k^{2}v_{\parallel}^{2}}\,\,\,,\label{15}\end{equation}
 and another arising from surface effects, \begin{equation}
P_{s}=e^{2}\omega_{p}^{4}\sum_{\mathbf{k}}\int d\omega\frac{1}{(\omega^{2}-\omega_{p}^{2}/2)(\omega^{2}-\omega_{p}^{2})}\cdot\frac{v_{\parallel}(i\mathbf{k}\mathbf{v_{\perp}}-kv_{\parallel})}{(\omega-\mathbf{k}\mathbf{v}_{\perp})^{2}+k^{2}v_{\parallel}^{2}}e^{-kv_{\parallel}t}e^{i(\mathbf{k}\mathbf{v}_{\perp}-\omega)t}\,\,.\label{16}\end{equation}
In performing the $\omega$-integrations in equations (\ref{15})
and (\ref{16}) we retain only the plasmon contributions arising from
the poles $\omega=\omega_{p}$ and $\omega=\omega_{p}/\sqrt{2}$.
For normal incidence ($v_{\perp}=0$, $v_{\parallel}=v$) we get easily
the well-known bulk contribution $P_{b}=\left(-e^{2}\omega_{p}^{2}/v\right)\ln(vk_{0}/\omega_{p})$,
where $k_{0}$ is an upper cut-off (associated, as usually, with the
ionization energy, or with the inverse of the mean inter-particle
spacing, etc), and the surface contribution \begin{equation}
P_{s}=-\frac{e^{2}\omega_{p}}{vt}\left(\sqrt{2}\sin\omega_{p}t/\sqrt{2}-\sin\omega_{p}t\right)\,\,.\label{17}\end{equation}
 We can see in equation (\ref{17}) the oscillatory behaviour of the
stopping power arising from the surface effects in the transient regime
near the surface.

\section{Interaction with the electromagnetic field. Polaritons}

We assume a plane wave incident on the plasma surface under angle
$\alpha$. Its frequency is given by $\omega=cK$, where $c$ is the
velocity of light and the wavevector $\mathbf{K}=(\mathbf{k},\kappa)$
has the in-plane component $\mathbf{k}$ and the perpendicular-to-plane
component $\kappa$, such as $k=K\sin\alpha$ and $\kappa=K\cos\alpha$.
In addition, $\mathbf{k}=k(\cos\varphi,\sin\varphi)$. The electric
field is taken as $\mathbf{E}_{0}=E_{0}(\cos\beta,0,-\sin\beta)e^{i\mathbf{kr}}e^{i\kappa z}e^{-i\omega t}$,
and we impose the condition $\cos\beta\sin\alpha\cos\varphi-\sin\beta\cos\alpha=0$
(transversality condition $\mathbf{K}\mathbf{E}_{0}=0$). The angle
$\beta$ defines the direction of the polarization of the incident
field. The geometry of the incident wave is shown in Fig. 1.%
\begin{figure}
\noindent \begin{centering}
\includegraphics{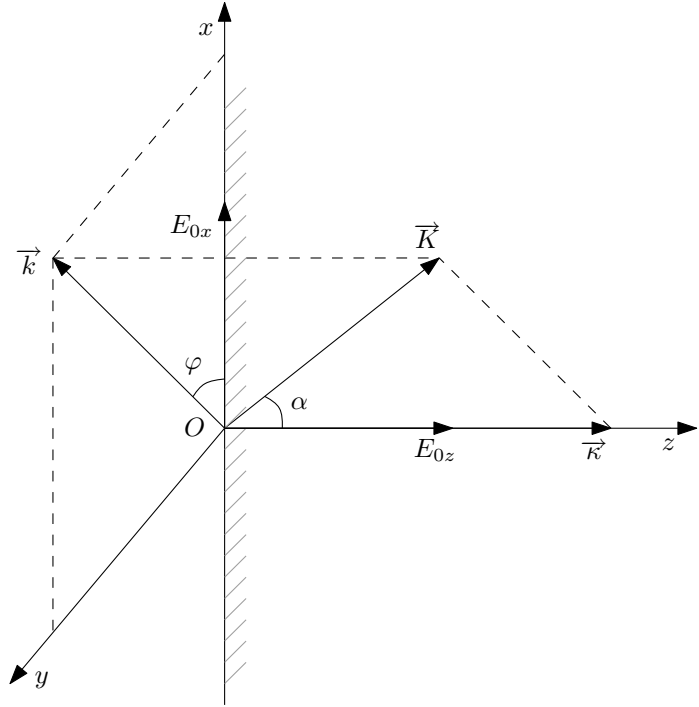}
\par\end{centering}

\caption{Electromagnetic plane wave $\mathbf{E}_{0}$, with wavevector $\mathbf{K}$,
incident on the surface $z=0$.}

\end{figure}

In the presence of an electromagnetic wave we use the equation of
motion \begin{equation}
\omega^{2}\mathbf{u}=\frac{e}{m}\mathbf{E}+\frac{e}{m}\mathbf{E}_{0}e^{i\kappa z}\,\,\,,\label{18}\end{equation}
for $z>0$, where $\mathbf{E}$ is the polarizing field; in equation
(\ref{18}) we have preseved explicitly only the $z$-dependence (\emph{i.e.}
we leave aside the factors $e^{i\mathbf{kr}}e^{-i\omega t}$). We
find it convenient to employ the vector potential \begin{equation}
\mathbf{A}(\mathbf{r},z;t)=\frac{1}{c}\int d\mathbf{r}'\int dz'\frac{\mathbf{j}(\mathbf{r}',z';t-R/c)}{R}\label{19}\end{equation}
 and the scalar potential \begin{equation}
\Phi(\mathbf{r},z;t)=\int d\mathbf{r}'\int dz'\frac{\rho(\mathbf{r}',z';t-R/c)}{R}\,\,\,,\label{20}\end{equation}

where \emph{$\mathbf{j}=-ne\dot{\mathbf{u}}\theta(z)e^{i\mathbf{kr}}e^{-i\omega t}$}
is the current density, $\rho=nediv\mathbf{u}=ne\left(i\mathbf{kv}+\frac{\partial u_{3}}{\partial z}\right)\theta(z)e^{i\mathbf{kr}}e^{-i\omega t}+neu_{3}(0)\delta(z)e^{i\mathbf{kr}}e^{-i\omega t}$
is the charge density and $R=\sqrt{(\mathbf{r}-\mathbf{r}')^{2}+(z-z')^{2}}$.
The integrals in equations (\ref{19}) and (\ref{20}) implies the
known integral\citep{key-33} \begin{equation}
\int_{\left|z\right|}^{\infty}dxJ_{0}\left(k\sqrt{x^{2}-z^{2}}\right)e^{i\omega x/c}=\frac{i}{\kappa}e^{i\kappa\left|z\right|}\,\,\,,\label{21}\end{equation}
where $J_{0}$ is the zeroth-order Bessel function of the first kind
(and $\omega^{2}/c^{2}=\kappa^{2}+k^{2}$). It is convenient to use
the projections of the in-plane displacement field $\mathbf{v}$ on
the vectors $\mathbf{k}$ and $\mathbf{k}_{\perp}=k(-\sin\varphi,\cos\varphi)$,
$\mathbf{k}_{\perp}\mathbf{k}=0$. We denote these components by $v_{1}=\mathbf{kv}/k$
and $v_{2}=\mathbf{k}_{\perp}\mathbf{v}/k$, and use also the components
$E_{1}=\mathbf{kE}/k$, $E_{2}=\mathbf{k}_{\perp}\mathbf{E}/k$ and
similar ones for the external field $\mathbf{E}_{0}$. We give here
the components of the external field \begin{equation}
E_{01}=E_{0}\cos\beta\cos\varphi\,\,,\,\, E_{02}=-E_{0}\cos\beta\sin\varphi\,\,,\,\, E_{03}=-E_{0}\sin\beta\,\,.\label{22}\end{equation}
 One can check immediately the transversality condition $E_{01}k+E_{03}\kappa=0$.
Making use of $\mathbf{E}=-\frac{1}{c}\frac{\partial\mathbf{A}}{\partial t}-grad\Phi$,
equations (\ref{19}) and (\ref{20}) give the electric field \begin{equation}
\begin{array}{c}
E_{1}=-2\pi ine\kappa\int_{0}dz'v_{1}(z')e^{i\kappa\left|z-z'\right|}-2\pi ne\frac{k}{\kappa}\int_{0}dz'u_{3}(z')\frac{\partial}{\partial z'}e^{i\kappa\left|z-z'\right|}\\
\\E_{2}=-2\pi ine\frac{\omega^{2}}{c^{2}\kappa}\int_{0}dz'v_{2}(z')e^{i\kappa\left|z-z'\right|}\\
\\E_{3}=2\pi ne\frac{k}{\kappa}\int_{0}dz'v_{1}(z')\frac{\partial}{\partial z}e^{i\kappa\left|z-z'\right|}-2\pi ine\frac{k^{2}}{\kappa}\int_{0}dz'u_{3}(z')e^{i\kappa\left|z-z'\right|}+4\pi neu_{3}\end{array}\label{23}\end{equation}
for $z>0$. It is worth observing in deriving these equations the
non-intervertibility of the derivatives and the integrals, according
to the identity \begin{equation}
\frac{\partial}{\partial z}\int_{0}dz^{'}f(z^{'})\frac{\partial}{\partial z^{'}}e^{i\kappa\left|z-z^{'}\right|}=\kappa^{2}\int_{0}dz^{'}f(z^{'})e^{i\kappa\left|z-z^{'}\right|}-2i\kappa f(z)\label{24}\end{equation}
 for any function $f(z)$, $z>0$; it is due to the discontinuity
in the derivative of the function $e^{i\kappa\left|z-z^{'}\right|}$
for $z=z^{'}$. Now, we employ equation of motion (\ref{18}) in equations
(\ref{23}) and get the integral equations \begin{equation}
\begin{array}{c}
\omega^{2}v_{1}=-\frac{i\omega_{p}^{2}\kappa}{2}\int_{0}dz'v_{1}(z')e^{i\kappa\left|z-z'\right|}-\frac{\omega_{p}^{2}k}{2\kappa}\int_{0}dz'u_{3}(z')\frac{\partial}{\partial z'}e^{i\kappa\left|z-z'\right|}+\frac{e}{m}E_{01}e^{i\kappa z}\\
\\\omega^{2}v_{2}=-\frac{i\omega_{p}^{2}\omega^{2}}{2c^{2}\kappa}\int_{0}dz'v_{2}(z')e^{i\kappa\left|z-z'\right|}+\frac{e}{m}E_{02}e^{i\kappa z}\\
\\\omega^{2}u_{3}=\frac{\omega_{p}^{2}k}{2\kappa}\int_{0}dz'v_{1}(z')\frac{\partial}{\partial z}e^{i\kappa\left|z-z'\right|}-\frac{i\omega_{p}^{2}k^{2}}{2\kappa}\int_{0}dz'u_{3}(z')e^{i\kappa\left|z-z'\right|}+\omega_{p}^{2}u_{3}+\frac{e}{m}E_{03}e^{i\kappa z}\end{array}\label{25}\end{equation}
for the coordinates $v_{1,2}$ and $u_{3}$ in the region $z>0$. 

The second equation (\ref{25}) can be solved straightforwardly by
noticing that \begin{equation}
\frac{\partial^{2}}{\partial z^{2}}\int_{0}dz'v_{2}(z')e^{i\kappa\left|z-z'\right|}=-\kappa^{2}\int_{0}dz'v_{2}(z')e^{i\kappa\left|z-z'\right|}+2i\kappa v_{2}\,\,.\label{26}\end{equation}
 We get \begin{equation}
\frac{\partial^{2}v_{2}}{\partial z^{2}}+(\kappa^{2}-\omega_{p}^{2}/c^{2})v_{2}=0\,\,.\label{27}\end{equation}
 The solution of this equation is \begin{equation}
v_{2}=\frac{2eE_{02}}{m\omega_{p}^{2}}\cdot\frac{\kappa\left(\kappa-\kappa^{'}\right)}{K^{2}}e^{i\kappa^{'}z}\,\,\,,\label{28}\end{equation}
where \begin{equation}
\kappa^{'}=\sqrt{\kappa^{2}-\omega_{p}^{2}/c^{2}}=\frac{1}{c}\sqrt{\omega^{2}\cos^{2}\alpha-\omega_{p}^{2}}\,\,.\label{29}\end{equation}
The wavevector $\kappa^{'}$can also be written in a more familiar
form $\kappa^{'}=(\omega/c)\sqrt{\varepsilon-\sin^{2}\alpha}$, where
$\varepsilon=1-\omega_{p}^{2}/\omega^{2}$ is the dielectric function.
The corresponding component of the (total) electric field (the refracted
field), can be obtained from equation (\ref{18}); it is given by
$\left(m\omega^{2}/e\right)v_{2}$. For $\kappa^{2}<\omega_{p}^{2}/c^{2}$
($\omega\cos\alpha<\omega_{p}$) this field does not propagate. For
$\kappa^{2}>\omega_{p}^{2}/c^{2}$ ($\omega$ greater than the transparency
edge $\omega_{p}/\cos\alpha$) it represents a refracted wave (transparency
regime) with the refraction angle $\alpha^{'}$ given by Snell's law
\begin{equation}
\frac{\sin\alpha^{'}}{\sin\alpha}=\frac{1}{\sqrt{1-\omega_{p}^{2}/\omega^{2}}}=1/\sqrt{\varepsilon}\,\,.\label{30}\end{equation}
The polariton frequency is given by \begin{equation}
\omega^{2}=c^{2}K^{2}=\omega_{p}^{2}+c^{2}K^{'2}\,\,\,,\label{31}\end{equation}
as it is well known, where $K^{'2}=\kappa^{'2}+k^{2}$. 

The first and the third equations (\ref{25}) can be solved by using
an equation similar with equation (\ref{26}) and by noticing that
they imply \begin{equation}
\kappa^{'2}u_{3}=ik\frac{\partial v_{1}}{\partial z}\,\,.\label{32}\end{equation}
We get \begin{equation}
v_{1}=\frac{2eE_{01}}{m\omega_{p}^{2}}\cdot\frac{\kappa^{'}\left(\kappa-\kappa^{'}\right)}{\kappa\kappa^{'}+k^{2}}e^{i\kappa^{'}z}\label{33}\end{equation}
 and \begin{equation}
u_{3}=\frac{2eE_{03}}{m\omega_{p}^{2}}\cdot\frac{\kappa\left(\kappa-\kappa^{'}\right)}{\kappa\kappa^{'}+k^{2}}e^{i\kappa^{'}z}\,\,.\label{34}\end{equation}
Similarly, the corresponding components of the refracted field are
given by equation (\ref{18}). It is easy to check the transversality
condition $v_{1}k+u_{3}\kappa^{'}=0$ (and the vanishing of the bulk
charge $ne\left(i\mathbf{kv}+\frac{\partial u_{3}}{\partial z}\right)=0$).%
\begin{figure}
\noindent \begin{centering}
\includegraphics[clip,scale=0.5]{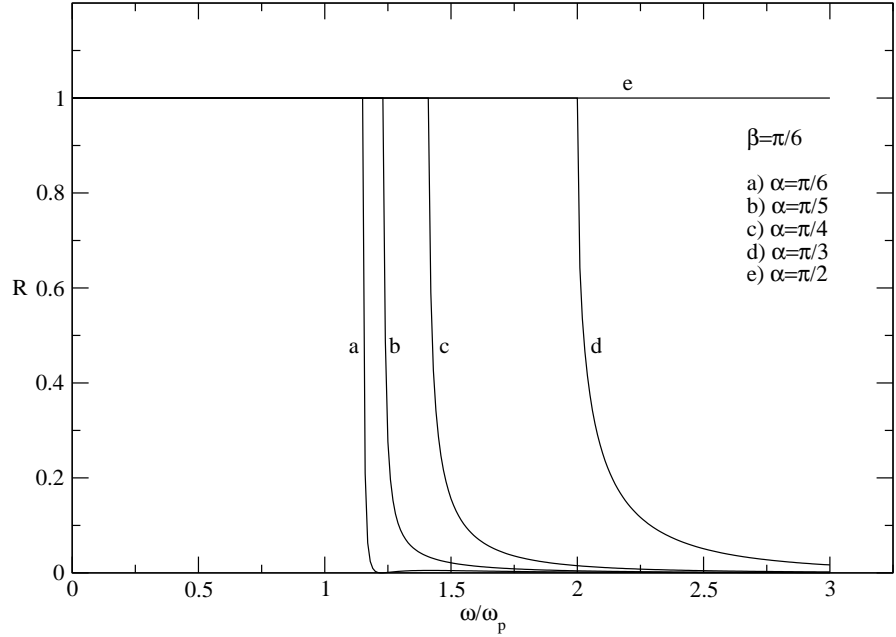}
\par\end{centering}

\caption{Reflection coefficient for a semi-infinite plasma for $\beta=\pi/6$
and various incidence angles $\alpha$. One can see the shoulder occurring
at the transparency edge $\omega_{p}/\cos\alpha$ and the zero occurring
at $\omega^{2}=\omega_{p}^{2}/\left(1-\tan^{2}\alpha\right)$ for
$\alpha=\beta=\pi/6$ ($R_{2}=0,$$\varphi=0$).}

\end{figure}

We can see that the polarization field $\mathbf{E}$ in equation (\ref{18})
cancels out the original incident field $\mathbf{E}_{0}$ and gives
the total, refracted field $m\omega^{2}\mathbf{u}/e$ inside the plasma.
This is an illustration of the so-called Ewald-Oseen extinction theorem.\citep{key-17,key-34} 

It is worth investigating the eigenvalues of the homogeneous system
of integral equations (\ref{25}), for parameter $\kappa$ given by
$\kappa=\sqrt{\omega^{2}/c^{2}-k^{2}}$. Such eigenvalues are given
by the roots of the vanishing denominator in equations (\ref{33})
and (\ref{34}), \emph{i.e.} by equation $\kappa\kappa^{'}+k^{2}=0$.
This equation has real roots for $\omega$ only for the damped regime,
\emph{i.e.} for $\kappa=i\left|\kappa\right|$ and $\kappa^{'}=i\left|\kappa^{'}\right|$.
Providing these conditions are satisfied, there is only one acceptable
branch of excitations, given by \begin{equation}
\omega^{2}=\frac{2\omega_{p}^{2}c^{2}k^{2}}{\omega_{p}^{2}+2c^{2}k^{2}+\sqrt{\omega_{p}^{4}+4c^{4}k^{4}}}\,\,.\label{35}\end{equation}
We can see that $\omega\sim ck$ in the long wavelength limit and
it approaches the surface-plasmon frequency $\omega\sim\omega_{p}/\sqrt{2}$
in the non-retarded limit ($ck\rightarrow\infty$). These excitations
are surface plasmon-polariton modes. We note that they imply $v_{2}=0$
and $v_{1},u_{3}\sim e^{-\left|\kappa^{'}\right|z}$. In addition,
a careful analysis of the homogeneous system of equations (\ref{25})
reveals another branch of excitations, given by $\omega=\omega_{p}$,
which, occurring in this context, may be termed the bulk plasmon-polariton
modes. They are characterized by $v_{2}=0$ and $v_{1}(\mathbf{k},0)=0$.
For all these modes we have $u_{3}=\left[ic^{2}k/\left(\omega^{2}-c^{2}k^{2}-\omega_{p}^{2}\right)\right]\frac{\partial v_{1}}{\partial z}$.

In order to get the reflected wave (the region $z<0$) we turn to
equations (\ref{23}) and use therein the solutions given above for
$v_{1,2}$ and $u_{3}$. It is worth noting here that the discontinuity
term $\omega_{p}^{2}u_{3}$ does not appear anymore in these equations
(because $z^{'}>0$ and $z<0$ and we cannot have $z=z^{'}$). The
integrations in equations (\ref{23}) are straightforward and we get
the field \begin{equation}
E_{1}=E_{01}\frac{\kappa-\kappa^{'}}{\kappa+\kappa^{'}}\cdot\frac{\kappa\kappa^{'}-k^{2}}{\kappa\kappa^{'}+k^{2}}e^{-i\kappa z}\,\,\,,\label{36}\end{equation}
 \begin{equation}
E_{2}=E_{02}\frac{\kappa-\kappa^{'}}{\kappa+\kappa^{'}}e^{-i\kappa z}\,\,\,\label{37}\end{equation}
and \begin{equation}
E_{3}=-E_{03}\frac{\kappa-\kappa^{'}}{\kappa+\kappa^{'}}\cdot\frac{\kappa\kappa^{'}-k^{2}}{\kappa\kappa^{'}+k^{2}}e^{-i\kappa z}\,\,.\label{38}\end{equation}
We can see that this field represents the reflected wave ($\kappa\rightarrow-\kappa$),
and we can check its transversality to the propagation wavevector.
Making use of the reflected field $\mathbf{E}_{refl}$ given by equations
(\ref{36})-(\ref{38}) and the refracted field $\mathbf{E}_{refr}$
obtained from equations (\ref{18}) and (\ref{23}) ($\mathbf{E}_{refr}=\mathbf{E}+\mathbf{E}_{0}=m\omega^{2}\mathbf{u}/e$)
one can check the continuity of the electric field and electric displacement
at the surface ($z=0$) in the form $E_{1,2refl}+E_{01,2}=E_{1,2refr}$,
$E_{3refl}+E_{03}=\varepsilon E_{3refr}$, where $\varepsilon=1-\omega_{p}^{2}/\omega^{2}$.
The angle of total polarization (Brewster's angle) is given by $\kappa\kappa^{'}-k^{2}=0$,
or $\tan^{2}\alpha=1-\omega_{p}^{2}/\omega^{2}=\varepsilon$ (for
$\alpha<\pi/4$). The above equations provide generalized Fresnel's
relations between the amplitudes of the reflected, refracted and incident
waves at the surface for any incidence angle and polarization. They
can also be written by using $\omega^{2}=\omega_{p}^{2}/\left(1-\varepsilon\right)$,
where $\varepsilon$ is the dielectric function.  

The reflection coefficient $R=\left|\mathbf{E}_{refl}\right|^{2}/\left|\mathbf{E}_{0}\right|^{2}$
can be obtained straightforwardly from the reflected fields given
by equations (\ref{36})-(\ref{38}). It can be written as \begin{equation}
R=R_{1}\left[\cos^{2}\beta\sin^{2}\varphi+R_{2}\left(\cos^{2}\beta\cos^{2}\varphi+\sin^{2}\beta\right)\right]\,\,\,,\label{39}\end{equation}
 where \begin{equation}
R_{1}=\left|\frac{\sqrt{\omega^{2}\cos^{2}\alpha-\omega_{p}^{2}}-\omega\cos\alpha}{\sqrt{\omega^{2}\cos^{2}\alpha-\omega_{p}^{2}}+\omega\cos\alpha}\right|^{2}\label{40}\end{equation}
 and \begin{equation}
R_{2}=\left|\frac{\cos\alpha\sqrt{\omega^{2}\cos^{2}\alpha-\omega_{p}^{2}}-\omega\sin^{2}\alpha}{\cos\alpha\sqrt{\omega^{2}\cos^{2}\alpha-\omega_{p}^{2}}+\omega\sin^{2}\alpha}\right|^{2}\,\,.\label{41}\end{equation}

The first term in the \emph{rhs} of equation (\ref{39}) corresponds
to $\beta=0$ ($\varphi=\pi/2$; $s$-wave, electric field perpendicular
to the plane of incidence, while the second term corresponds to $\beta=\alpha$
($\varphi=0$; $p$-wave, electric field in the plane of incidence).
It is easy to see that there exists a cusp (shoulder) in the behaviour
of the function $R(\omega)$, occurring at the transparency edge $\omega=\omega_{p}/\cos\alpha$,
where the reflection coefficient exhibits a sudden enhancement on
passing from the propagating regime to the damped one, as expected
(total reflection). The condition for total reflection can also be
written as $\sin\alpha=\sqrt{\varepsilon}$, where $R=1$ ($R_{1,2}=1$),
as it is well known. For illustration, the reflection coefficient
is shown in Fig. 2 for $\beta=\pi/6$ and various incidence angles.
The reflection coefficient is vanishing at $\omega^{2}=\omega_{p}^{2}/\left(1-\tan^{2}\alpha\right)$
for $\alpha=\beta<\pi/4$ ($R_{2}=0,\varphi=0$). %
\begin{figure}
\noindent \begin{centering}
\includegraphics[clip,scale=0.5]{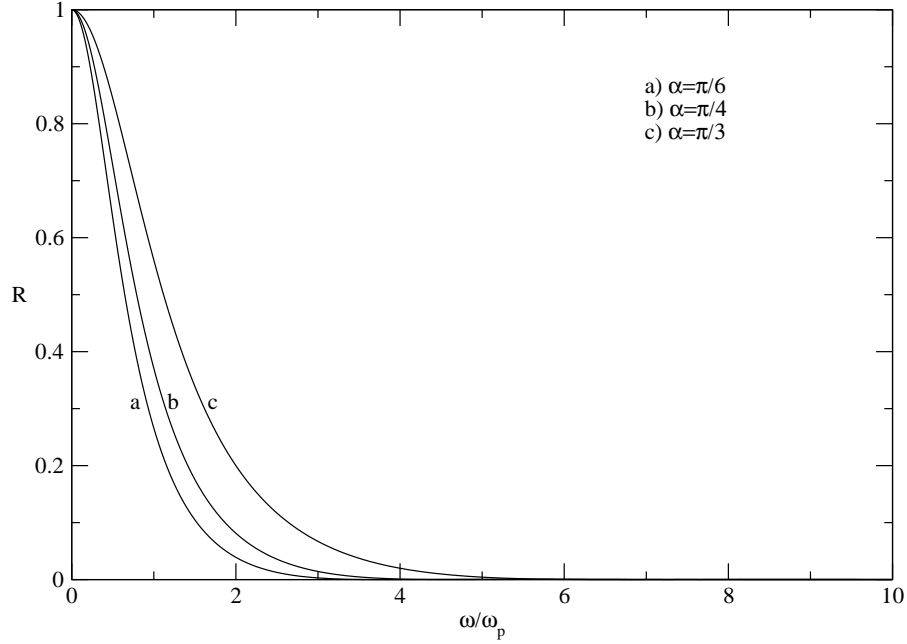}
\par\end{centering}

\caption{Reflection coefficient for a slab of thickness $d$ ($d\omega_{p}/c=1$)
for $\beta=0$, $\varphi=\pi/2$ ($s$-wave) and a few incidence angles
$\alpha$. Its slope is continuous at the transparency edge ($\omega\cos\alpha=\omega_{p}$).
The oscillations occurring in the transparency regime are too small
to be visible in Figure.}

\end{figure}

\section{Plasma slab}

We consider an ideal plasma slab of thickness $d$, extending over
the region $0<z<d$ and bounded by the vacuum. The displacement field
$\mathbf{u}$ can be represented as $(\mathbf{v},u_{3})\left[\theta(z)-\theta(z-d)\right]$,
where $\mathbf{v}$ is the displacement component in the $(x,y)$-plane
and $u_{3}$ is the displacement component along the $z$-direction.
The approach presented above for a semi-infinite plasma can easily
be extended to this case. The analogous of the equation of motion
(\ref{4}) exhibits now two polarization contributions, arising from
the two surfaces. The dielectric response similar to equation (\ref{9})
is given by \begin{equation}
\begin{array}{c}
\mathbf{v}=\frac{ie\mathbf{k}\omega_{p}^{2}}{m}\cdot\frac{\left(2\omega^{2}-\omega_{p}^{2}\right)\Phi_{0}-\omega_{p}^{2}\Phi_{d}e^{-kd}}{\left(\omega^{2}-\omega_{p}^{2}\right)\left[2\omega^{2}-\omega_{p}^{2}\left(1-e^{-kd}\right)\right]\left[2\omega^{2}-\omega_{p}^{2}\left(1+e^{-kd}\right)\right]}e^{-kz}+\\
\\+\frac{ie\mathbf{k}\omega_{p}^{2}}{m}\cdot\frac{\left(2\omega^{2}-\omega_{p}^{2}\right)\Phi_{d}-\omega_{p}^{2}\Phi_{0}e^{-kd}}{\left(\omega^{2}-\omega_{p}^{2}\right)\left[2\omega^{2}-\omega_{p}^{2}\left(1-e^{-kd}\right)\right]\left[2\omega^{2}-\omega_{p}^{2}\left(1+e^{-kd}\right)\right]}e^{kz-kd}-\frac{ie\mathbf{k}}{m}\frac{\Phi}{\omega^{2}-\omega_{p}^{2}}\end{array}\label{42}\end{equation}
 and $iku_{3}=\frac{\partial v}{\partial z}$, where $\Phi_{0}=\Phi(z=0)$,
$\Phi_{d}=\Phi(z=d)$, $0<z<d$. The electric field is given by $E_{\perp}=4\pi nev$
and $E_{\parallel}=4\pi neu_{3}$. One can see that, beside the bulk
plasmon mode $\omega_{p}^{2}$, there appears two surface modes given
by $\omega_{p}^{2}\left(1\pm e^{-kd}\right)/2$, as it is well known.
For $d\rightarrow\infty$ equation (\ref{42}) becomes the first equation
(\ref{9}) for the semi-infinite plasma. For $d\rightarrow0$ we get
the well-known plasma frequency $\sqrt{\left(2\pi n_{s}e^{2}/m\right)k}$
for a sheet with surface electron density $n_{s}=nd$. %
\begin{figure}
\noindent \begin{centering}
\includegraphics[clip,scale=0.5]{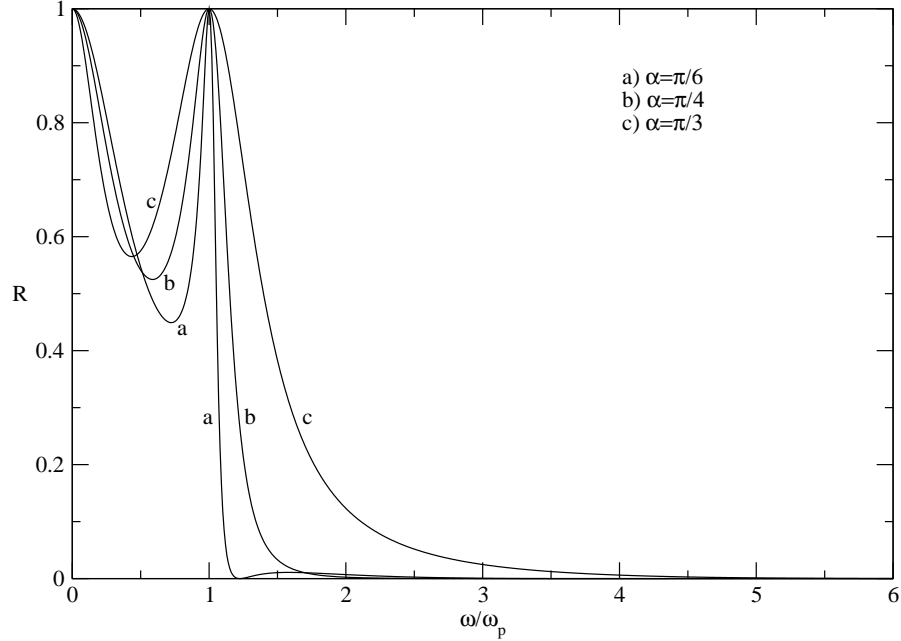}
\par\end{centering}

\caption{Reflection coefficient for a slab of thickness $d$ ($d\omega_{p}/c=1$)
for $\alpha=\beta$, $\varphi=0$ ($p$-wave) and a few incidence
angles $\alpha$. It exhibits a local maximum ($R=1$) for $\omega=\omega_{p}$
and small oscillations in the transparency region $\omega\cos\alpha>\omega_{p}$
(too small to be visible in Figure). In addition, it is vanishing
for $\omega^{2}=\omega_{p}^{2}/\left(1-\tan^{2}\alpha\right)$, $\alpha<\pi/4$,
as one can see in Figure for $\alpha=\pi/6$ (curve $a$). }

\end{figure}

The bulk contribution to the energy loss is the same as for the semi-infinite
plasma. We compute the surface contrbution to the electron energy
loss for $kd\gg\omega_{p}d/v\gg1$, \emph{i.e.} for a fast electron
moving with velocity $v$, which, however, spends enough time in the
sample to excite plasmons. For normal incidence the surface contribution
consists of two oscillatory terms \begin{equation}
\begin{array}{c}
P_{s}=-\frac{e^{2}\omega_{p}}{vt}\left(\sqrt{2}\sin\omega_{p}t/\sqrt{2}-\sin\omega_{p}t\right)-\\
\\-\frac{e^{2}\omega_{p}}{d-vt}\left[\sqrt{2}\sin\omega_{p}\left(d/v-t\right)/\sqrt{2}-\sin\omega_{p}\left(d/v-t\right)\right]\,\,\,,\end{array}\label{43}\end{equation}
 corresponding to the two surfaces, for $0<t<d/v$. The total energy
loss during the passage through the slab is given by \begin{equation}
\int_{0}^{d/v}dtP_{s}\simeq\int_{0}^{\infty}dtP_{s}=-\pi\left(\sqrt{2}-1\right)\frac{e^{2}\omega_{p}}{v}\,\,.\label{44}\end{equation}

We use again the equation of motion (\ref{18}) and the retarded potentials
given by equations (\ref{19}) and (\ref{20}) in order to get the
refracted field (field inside the slab), reflected ($z<0$) and transmitted
($z>d$) fields. The polarization field is given by the same equations
(\ref{23}), where the $z$-integration is limited to the region $0<z<d$.
The same holds for the equations of motion (\ref{25}). We solve these
equations by the same method used above. Within the slab we have two
waves of the form $e^{\pm i\kappa^{'}z}$, one being the refracted
wave through the first surface ($z=0$), the other being the reflected
wave on the second surface ($z=d$). The wavevector $\kappa^{'}$
is given by the same equation (\ref{29}), and the transparency edge
is given by the same condition $\omega\cos\alpha=\omega_{p}$ as for
a semi-infinite plasma. We get \begin{equation}
v_{2}=A_{2}\left[e^{i\kappa^{'}z}-\frac{\kappa-\kappa^{'}}{\kappa+\kappa^{'}}e^{2i\kappa^{'}d}\cdot e^{-i\kappa^{'}z}\right]\,\,\,,\label{45}\end{equation}
 where \begin{equation}
A_{2}=\frac{2eE_{02}}{m\omega_{p}^{2}}\cdot\frac{\kappa\left(\kappa-\kappa^{'}\right)\left(\kappa+\kappa^{'}\right)^{2}}{K^{2}\left[\left(\kappa+\kappa^{'}\right)^{2}-\left(\kappa-\kappa^{'}\right)^{2}e^{2i\kappa^{'}d}\right]}\,\,\,,\label{46}\end{equation}
 and \begin{equation}
v_{1}=A_{1}\left[e^{i\kappa^{'}z}-\frac{\kappa-\kappa^{'}}{\kappa+\kappa^{'}}\cdot\frac{\kappa\kappa^{'}-k^{2}}{\kappa\kappa^{'}+k^{2}}e^{2i\kappa^{'}d}\cdot e^{-i\kappa^{'}z}\right]\,\,\,,\label{47}\end{equation}
 where \begin{equation}
A_{1}=\frac{2eE_{01}}{m\omega_{p}^{2}}\cdot\frac{\kappa^{'}\left(\kappa-\kappa^{'}\right)\left(\kappa+\kappa^{'}\right)^{2}\left(\kappa\kappa^{'}+k^{2}\right)}{\left(\kappa+\kappa^{'}\right)^{2}\left(\kappa\kappa^{'}+k^{2}\right)^{2}-\left(\kappa-\kappa^{'}\right)^{2}\left(\kappa\kappa^{'}-k^{2}\right)^{2}e^{2i\kappa^{'}d}}\,\,;\label{48}\end{equation}
 the third component can be obtained from $\kappa^{'2}u_{3}=ik\left(\partial v_{1}/\partial z\right)$.
One can check the transversality of these waves and can compute the
dispersion relations for the eigenvalues (bulk and surface plasmon-polaritons)
in the like manner as for the semi-infinite plasma. %
\begin{figure}
\noindent \begin{centering}
\includegraphics[clip,scale=0.5]{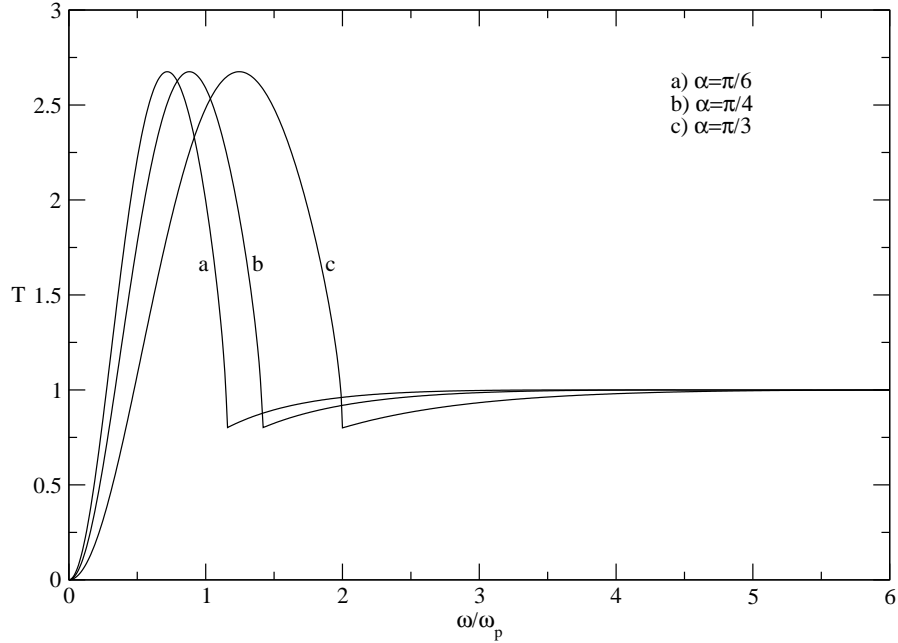}
\par\end{centering}

\caption{Transmission coefficient for a slab of thickness $d$ ($d\omega_{p}/c=1$)
for $\beta=0$, $\varphi=\pi/2$ ($s$-wave) and a few incidence angles
$\alpha$. One can see the characteristic cusp at the transparency
edge $\omega\cos\alpha=\omega_{p}$ and the peak ocurring below this
edge. The oscillations occurring in the transparency regime are too
small to be visible in Figure.}

\end{figure}

The reflected field is given by \begin{equation}
\begin{array}{c}
E_{1}=E_{01}\left(1-e^{2i\kappa^{'}d}\right)\frac{\left(\kappa^{2}-\kappa^{'2}\right)\left(\kappa^{2}\kappa^{'2}-k^{4}\right)}{\left(\kappa+\kappa^{'}\right)^{2}\left(\kappa\kappa^{'}+k^{2}\right)^{2}-\left(\kappa-\kappa^{'}\right)^{2}\left(\kappa\kappa^{'}-k^{2}\right)^{2}e^{2i\kappa^{'}d}}e^{-i\kappa z}\,\,\,,\\
\\E_{2}=E_{02}\left(1-e^{2i\kappa^{'}d}\right)\frac{\kappa^{2}-\kappa^{'2}}{\left(\kappa+\kappa^{'}\right)^{2}-\left(\kappa-\kappa^{'}\right)^{2}e^{2i\kappa^{'}d}}e^{-i\kappa z}\,\,\,\end{array}\label{49}\end{equation}
 and $E_{3}=-E_{03}\left(E_{1}/E_{01}\right)$.

>From the above results one can check the continuity of the electric
field and electric displacement as well as the angle of total polarization
given by $\tan^{2}\alpha=1-\omega_{p}^{2}/\omega^{2}=\varepsilon$.
If we take formally $e^{2i\kappa^{'}d}\rightarrow0$ we recover all
the fields for the semi-infinite plasma. Indeed, for the semi-infinite
plasma all the integrations to $z\rightarrow\infty$ are taken by
assuming a vanishing factor $e^{-\mu z}$, $\mu>0$, and letting $\mu$
go to zero. If we preserve this factor for the slab, it gives rise
to factors of the form $e^{2i\kappa^{'}d}e^{-\mu d}$, which are vanishing
for $d\rightarrow\infty$. The limit $d\rightarrow0$ (plasma sheet)
cannot be taken directly on the above results ($\omega_{p}\sim1/\sqrt{d}$,
$\kappa^{'}\sim i\omega_{p}/c$), because of the discontinuities arising
from the $\theta$-function. The calculations for a plasma sheet with
a finite (superficial) charge density $n_{s}$ must be done separately.
They are left, together with other related results, for a forthcoming
publication. The limit $\kappa^{'}d\ll1$ ($\kappa d\ll1$) can be
taken directly on the formulae given here. It corresponds to wavelengths
much longer than the thickness of the slab. %
\begin{figure}
\noindent \begin{centering}
\includegraphics[clip,scale=0.5]{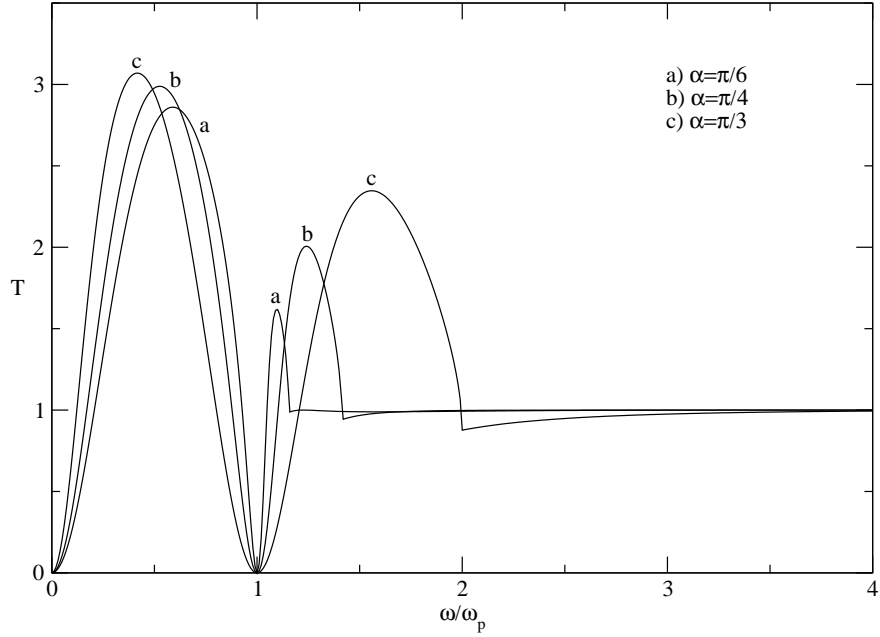}
\par\end{centering}

\caption{Transmission coefficient for a slab of thickness $d$ ($d\omega_{p}/c=1$)
for a few incidence angles $\alpha=\beta$ and $\varphi=0$ ($p$-wave).
One can see the two peaks occurring below the transparency edge $\omega\cos\alpha=\omega_{p}$
(the cusp in Figure) and the zero for $\omega=\omega_{p}$. The oscillations
occurring in the transparency regime are too small to be visible in
Figure.}

\end{figure}

The reflection coefficient for the plasma slab $R=\left|\mathbf{E}_{refl}\right|^{2}/\left|\mathbf{E}_{0}\right|^{2}$,
where the reflected field is given by equations (\ref{49}), has a
different structure than the reflection coefficient for the semi-infinite
plasma. It can be written as \begin{equation}
R=\frac{\omega_{p}^{4}}{c^{4}}\left|1-e^{2i\kappa^{'}d}\right|^{2}\left[R_{1}\cos^{2}\beta\sin^{2}\varphi+R_{2}\left(\cos^{2}\beta\cos^{2}\varphi+\sin^{2}\beta\right)\right]\,\,\,,\label{50}\end{equation}
 where \begin{equation}
R_{1}=\frac{1}{\left|\left(\kappa+\kappa^{'}\right)^{2}-\left(\kappa-\kappa^{'}\right)^{2}e^{2i\kappa^{'}d}\right|^{2}}\,\,\,\label{51}\end{equation}
 and \begin{equation}
R_{2}=\frac{\left|\kappa^{2}\kappa^{'2}-k^{4}\right|^{2}}{\left|\left(\kappa+\kappa^{'}\right)^{2}\left(\kappa\kappa^{'}+k^{2}\right)^{2}-\left(\kappa-\kappa^{'}\right)^{2}\left(\kappa\kappa^{'}-k^{2}\right)^{2}e^{2i\kappa^{'}d}\right|^{2}}\,\,.\label{52}\end{equation}
The reflection coefficient given by equation (\ref{50}) is shown
in Figs. 3-4 for $\beta=0$, $\varphi=\pi/2$ ($s$-wave) and, respectively,
$\alpha=\beta$, $\varphi=0$ ($p$-wave) and $d\omega_{p}/c=1$.
The reflection coefficient exhibits characteristic oscillations arising
from the exponential factor in equations (\ref{50})-(\ref{52}) and
has an abrupt enhancement in the damping regime. In addition, $R_{2}$
is vanishing for $\omega^{2}=\omega_{p}^{2}/\left(1-\tan^{2}\alpha\right)$
($\alpha<\pi/4$) and $R_{2}=1$ for $\omega=\omega_{p}$. 

The transmitted field (region $z>d$) is given by \begin{equation}
\begin{array}{c}
E_{1}=E_{01}\frac{4K^{2}\kappa\kappa^{'}\left(\kappa^{'2}+k^{2}\right)e^{i\left(\kappa^{'}-\kappa\right)d}}{\left(\kappa+\kappa^{'}\right)^{2}\left(\kappa\kappa^{'}+k^{2}\right)^{2}-\left(\kappa-\kappa^{'}\right)^{2}\left(\kappa\kappa^{'}-k^{2}\right)^{2}e^{2i\kappa^{'}d}}e^{i\kappa z}\\
\\E_{2}=E_{02}\frac{4\kappa^{'}\kappa e^{i\left(\kappa^{'}-\kappa\right)d}}{\left(\kappa+\kappa^{'}\right)^{2}-\left(\kappa-\kappa^{'}\right)^{2}e^{2i\kappa^{'}d}}e^{i\kappa z}\end{array}\label{53}\end{equation}
 and $E_{3}=E_{03}\left(E_{1}/E_{01}\right)$. One can check the continuity
of the electric field and electric displacement at the surface $z=d$.
In the limit $d\rightarrow\infty$ the transmitted field is vanishing.
The transmission coefficient given by $T=\left|\mathbf{E}_{tr}\right|^{2}/\left|\mathbf{E}_{0}\right|^{2}$,
where $\mathbf{E}_{tr}$ is given by equations (\ref{53}), can be
written as \begin{equation}
T=16\kappa^{2}\left|\kappa^{'}\right|^{2}\left[R_{1}\cos^{2}\beta\sin^{2}\varphi+\frac{K^{4}\left|\kappa^{'2}+k^{2}\right|^{2}}{\left|\kappa^{2}\kappa^{'2}-k^{4}\right|^{2}}R_{2}\left(\cos^{2}\beta\cos^{2}\varphi+\sin^{2}\beta\right)\right]\,\,\,,\label{54}\end{equation}
 where $R_{1,2}$ are given by equations (\ref{51}) and (\ref{52}).
This transmission coefficient is shown in Figs. 5-6 for $\beta=0$,
$\varphi=\pi/2$ ($s$-wave) and, respectively, $\alpha=\beta$, $\varphi=0$
($p$-wave) and $d\omega_{p}/c=1$. Beside the characteristic cusp
occurring at the transparency edge ($\omega\cos\alpha=\omega_{p}$),
the transmission coefficient exhibits an appreciable enhancement below
this edge. For $\alpha=\beta$, $\varphi=0$ ($p$-wave) and $\omega=\omega_{p}$
the reflection coefficient attains the value unity and the transmission
coefficient vanishes. The fields derived above can be viewed as generalized
Fresnel's relations for a plasma slab.

\section{Conclusions}

The approach presented here is a quasi-classical one, valid for wavelengths
much longer than the amplitude of the Fourier components of the displacement
field $\mathbf{u}$. This is not a particularly restrictive condition
for the classical dynamics of the electromagnetic field interacting
with matter. When this condition is violated, as, for instance, for
wavelengths much shorter than the mean separation distance between
electrons, there appear both higher-order terms in the equations of
motion and the coupling to the individual motion of the electrons.
These couplings affect in general the dispersion relations and introduce
a finite lifetime (damping) for the plasmon and polariton modes. 

Making use of the equations of motion for the displacement field $\mathbf{u}$
and the radiation formulae for the electromagnetic potentials, we
have computed herein the plasmon and polariton modes for an ideal
semi-infinite electron plasma and an ideal plasma slab of finite thickness,
as well as the dielectric response, the electron energy loss, the
reflected and refracted waves and the reflection coefficient. For
the semi-infinite plasma we have identified the bulk and surface plasmon-polariton
modes and for the plasma slab we have computed also the transmitted
wave and the transmission coefficient. It was shown that the stopping
power due to the surface effects has a characteristic oscillatory
behaviour in the transient regime near the surfaces. The field inside
the plasma is either damped (evanescent) or propagating, as it is
well known, and the reflection coefficient for the semi-infinite plasma
exhibits a sudden enhancement on passing from the propagating to the
damped regime, as expected. The transparency edge is given by $\omega\cos\alpha=\omega_{p}$,
where $\alpha$ is the incidence angle, $\omega$ is the frequency
of the incident wave and $\omega_{p}$ is the plasma frequency. Apart
from characteristic oscillations, the reflection and transmission
coefficients for the plasma slab exhibit an appreciable enhancement
below the transparency edge. 

Other effects related to the dynamics of a semi-infinite electron
plasma, or, in general, various plasmas with rectangular geometries,
can be computed similarly by using the method presented here. The
method can also be applied to plasmas with other, more particular,
geometries. The dissipation can be introduced (as for metals) and
a model can be formulated for dielectrics, amenable to the method
presented here. This will allow the treatment of more realistic cases
as well as various interfaces, in particular plasmas (or metals) bounded
by dielectrics. These investigations are left for forthcoming publications.

\textbf{Acknowledgments.} The authors are indebted to the members
of the Laboratory of Theoretical Physics at Magurele-Bucharest for
many useful discusssions, and to dr. L. C. Cune for his help in various
stages of this work.

\end{document}